\documentclass[pra,reprint,preprintnumbers,
amsmath,amssymb, showkeys
]{revtex4-2}

\usepackage{graphicx}
\usepackage{dcolumn}
\usepackage{bm}
\usepackage{amssymb}
\usepackage{amsmath}
\usepackage{color}
\usepackage{soul}
\usepackage{subfig}
\usepackage{appendix}
\usepackage{booktabs}
\usepackage{comment}
\usepackage[hyperindex,breaklinks,hidelinks]{hyperref}

\newcommand{\id}{1\!\!1}

\begin{document}
	
	\title{Critical Spectrum and Quantum Criticality in the Two-Photon Rabi-Stark Model}
	
	\author{Jiong Li$^{1}$}
	\author{Qing-Hu Chen$^{1,2,}$}
	\email{qhchen@zju.edu.cn}
	
	\affiliation{$^{1}$Zhejiang Key Laboratory of Micro-Nano Quantum Chips and Quantum Control, School of Physics, Zhejiang University, Hangzhou 310027, China\\
		$^{2}$Collaborative Innovation Center of Advanced Microstructures, Nanjing University, Nanjing 210093, China}
	
	\date{\today}
	
	\begin{abstract}
		We investigate the spectral properties and quantum criticality of the two-photon Rabi-Stark model. Using the exact solution of this model, we rigorously derive a condition for complete spectral collapse, where all bound states vanish.  In this case, the energy gap closes at a critical coupling, signaling a continuous quantum phase transition. The corresponding gap exponent differs from those in both the one-photon Rabi-Stark model and the quantum Rabi model, suggesting a distinct universality class. While in the general case, an infinite number of discrete bound states exist when spectral collapse occur and the energy gap remains open. By mapping to an inverse square potential well,  these bound levels approach the threshold energy exponentially. Our results offer new insights into novel spectral phenomena in nonlinear quantum Rabi models, with potential implications for experimental realizations in circuit QED and trapped ion systems.
	\end{abstract}
	
	
	\maketitle
	
	\section{Introduction}
	\label{sec:intro}
	% TODO: write your article here.
	The quantum Rabi model (QRM) is widely recognized as the most fundamental framework for describing the interaction between a qubit (a two-level system) and a single quantized mode of the electromagnetic field \cite{rabi_process_1936, braak_semi-classical_2016}. It has been extensively applied in quantum optics \cite{scully_quantum_1997, meystre_elements_2007, meystre_quantum_2021}, quantum information \cite{ ashhab_qubit-oscillator_2010}, condensed matter physics \cite{Wagner:1986wp}, and remains a central focus of ongoing research.
	
	With the rapid advancement of quantum technologies, the regimes of strong \cite{wallraff_strong_2004, hennessy_quantum_2007}, ultrastrong \cite{peropadre_switchable_2010, forn-diaz_observation_2010, forn-diaz_ultrastrong_2019}, and even deep strong coupling \cite{casanova_deep_2010, yoshihara_superconducting_2017} have been experimentally realized across various platforms, including cavity quantum electrodynamics (QED) \cite{forn-diaz_ultrastrong_2017, forn-diaz_ultrastrong_2019}, circuit QED \cite{wallraff_strong_2004, peropadre_switchable_2010}, trapped ions \cite{leibfried_quantum_2003, pedernales_quantum_2015}, and quantum dots \cite{hennessy_quantum_2007, englund_controlling_2007}. In these regimes, higher order nonlinear terms become significant and must be taken into account. Consequently, considerable experimental and theoretical efforts have been devoted to various generalizations of the QRM, including two-photon interactions \cite{bertet_generating_2002, chen_exact_2012, cong_polaron_2019}, anisotropic coupling \cite{xie_anisotropic_2014}, Stark-like nonlinear interactions \cite{grimsmo_cavity-qed_2013, eckle_generalization_2017, xie_quantum_2019, braak2024c}.
	
	Certain extensions of the QRM exhibit an unconventional phenomenon known as spectral collapse. In the two-photon quantum Rabi model (tpQRM)—a natural generalization of the QRM—spectral collapse arises when the qubit–cavity coupling reaches a critical value, resulting in an energy spectrum that comprises both discrete and continuous components \cite{duan_two-photon_2016, chan_bound_2020, lo_manipulating_2020, rico_spectral_2020, braak_spectral_2023}. Although strictly continuous spectra cannot occur in confined systems—since higher order nonlinear terms must be incorporated into the Hamiltonian to maintain bound states—experimental findings suggest that extremely dense energy levels can emerge near the critical point in certain implementations \cite{felicetti_spectral_2015, felicetti_two-photon_2018, felicetti_ultrastrong-coupling_2018, piccione_two-photon-interaction_2022}. Recently, the tpQRM has also attracted considerable interest within the pure mathematics community \cite{nakahama_equivalence_2025, hiroshima_fiber_2025}.
	
	Furthermore, motivated by the excellent agreement between perturbative predictions of the QRM and the experimentally observed Bloch–Siegert frequency shift \cite{forn-diaz_observation_2010}, Grimsmo and Parkins \cite{grimsmo_cavity-qed_2013} introduced a dynamical Stark nonlinear term in quantum optics, representing the quantum analogue of the Bloch–Siegert shift. The resulting model was later termed the Rabi–Stark model (RSM). In contrast to the original QRM, which requires unphysical parameter regimes to induce a phase transition, the RSM enables the realization of a superradiant phase transition within a physically accessible parameter range \cite{grimsmo_open_2014}. Inspired by this finding, numerous studies have investigated the spectral properties of the RSM and confirmed the possibility of spectral collapse in this model \cite{eckle_generalization_2017, xie_quantum_2019, braak2024c}.
	
	The quantum phase transition (QPT) in the Dicke model in the thermodynamic limit has been known for a long time \cite{emary_quantum_2003, emary_chaos_2003}, and the QPT in the QRM in the limit of an infinite ratio between the qubit and cavity frequencies was discovered a decade ago \cite{ashhab_superradiance_2013, hwang_quantum_2015}, both belong to the same universality class. On the other hand, it has been observed that the two-photon Dicke model also undergoes a QPT \cite{garbe_superradiant_2017}, sharing the same universality class as the one-photon Dicke and Rabi models. However, the QPT in the tpQRM has not yet been reported in the literature, to the best of our knowledge.
	
	Recently, we derived a condition under which the energy spectrum becomes fully continuous during spectral collapse in the anisotropic tpQRM ~\cite{li_critical_2025}. Given its potential applications in critically enhanced quantum metrology \cite{ying2022}, achieving the critical coupling strength and anisotropic qubit-cavity coupling experimentally remains a significant challenge. The nonlinear Stark term introduced in the isotropic tpQRM—referred to as the two-photon Rabi–Stark model (tpRSM)—enables a first order phase transition at finite coupling strengths and preserves spectral collapse, but at comparatively weaker coupling strengths \cite{li_two-photon_2020}. This naturally raises the question: Does the tpRSM admit a critical condition under which the spectrum becomes fully continuous, thereby relaxing the experimental constraints on anisotropy and strong coupling strength? If so, could this lead to a deeper understanding of the associated QPT?
	
	The Hamiltonian of the tpRSM is given by
	\begin{equation}
		H = \omega a^\dagger a + g \sigma_z \left[ a^2 + (a^\dagger)^2 \right] - \sigma_{x} \left( \frac{\Delta}{2} + U a^\dagger a \right),
		\label{H_origin}
	\end{equation}
	where $a$ and $a^\dagger$ are the photon annihilation and creation operators for a single cavity mode with frequency $\omega$; $\Delta$ denotes the qubit energy splitting; $g$ is the qubit–cavity coupling strength; and $\{ \sigma_{x}, \sigma_{y}, \sigma_{z} \}$ are the Pauli matrices. For simplicity, we set $\omega = 1$ throughout this paper. By introducing the symmetry operator $\Pi = \sigma_{x} \otimes \exp \left[i \frac{\pi}{2} a^{\dagger}a \right]$, which satisfies $\Pi^4 = \id$, we find that $\Pi H \Pi^{\dagger} = H$, indicating that the Hamiltonian possesses a $\mathbb{Z}_4$ symmetry.
	
	The experimental realization of this model can be directly achieved using the circuit QED architecture proposed by Felicetti \textit{et al.} \cite{felicetti_two-photon_2018}, where the Stark term arises from an additional inductance induced by the qubit. Moreover, a proposal to realize the one-photon RSM in a trapped-ion experiment has been put forward \cite{cong_selective_2020}. In that scheme, an external laser field mediates the interaction between the electronic transitions and the motional degrees of freedom, thereby generating the Stark term. Since a proposal for the realization of the tpQRM in trapped-ion systems already exists \cite{felicetti_spectral_2015}, simulating the tpRSM in the same setup via laser driving appears to be experimentally feasible.
	
	\section{Energy spectra close to the critical coupling}
	
	The analytically exact solutions to the tpRSM have been provided in Ref.~\cite{li_two-photon_2020} using the Bogoliubov operator approach, which is briefly described in the Appendix for completeness. The derived confluent Heun function, also called $G$ function \cite{braak_integrability_2011}, is given by	
	\begin{equation}
		G_{\pm }^{(q)}(E)=\sum_{n=0}^{\infty }\left( e_{n}^{(q)}\mp
		f_{n}^{(q)}\right) \frac{\left[ 2(n+q-\frac{1}{4})\right] !}{2^{n}n!}\tanh
		^{n}\theta ,  \label{G_func}
	\end{equation}
	where the parameter $\theta $ is defined by $\tanh \theta =\sqrt{(1-\beta) / (1+\beta)}$ with $\beta = \sqrt{1-4\gamma^{2} g^{2}}$ and $\gamma^{-1} =  \sqrt{1-U^{2}}$. The coefficients $e_{n}^{(q)}$
	and $f_{n}^{(q)}$ are given by
	\begin{widetext}
			\begin{equation}
			e_{n}^{(q)} = \frac{\frac{\Delta}{2} + \frac{2U}{\beta}(n+q) - \frac{U}{2} - \frac{U\gamma}{\gamma+1} \left[ 2\frac{1+4\gamma g^{2}}{\beta}(n+q) - \frac{1}{2} - E \right]}{2\frac{1-4\gamma g^{2}}{\beta}(n+q) - \frac{1}{2} - E - \frac{U\gamma}{\gamma +1} \left[ \frac{\Delta}{2} + \frac{2U}{\beta}(n+q) - \frac{U}{2} \right]} f_{n}^{(q)} = \Omega_{n}^{(q)} f_{n}^{(q)},  \label{enq}
		\end{equation}
		\begin{equation}
			f_{n+1}^{(q)} = \frac{\left\{ 2(1+4\gamma g^{2})(n+q) - \beta \left( \frac{1}{2} + E \right) -\left[ \beta \frac{\Delta -U}{2} + 2U(n+q) \right] \Omega_{n}^{(q)} \right\} f_{n}^{(q)} - \left[ \gamma + 1 - U \gamma \Omega_{n-1}^{(q)} \right] g f_{n-1}^{(q)}}{4g \left( n+q+\frac{1}{4} \right) \left( n+q+\frac{3}{4} \right) \left[ \gamma + 1 - U \gamma \Omega_{n+1}^{(q)} \right] }.  \label{fnq}
		\end{equation}
	\end{widetext}
	Here, $f_{n}^{(q)}$ can be obtained from the three-term recurrence relation \eqref{fnq}, starting with $f_{0} \equiv 1$. The sign $\pm $ in the $G$ function \eqref{G_func} indicates even and odd parity respectively within each Bargmann subspace $\mathcal{H}_{q}$, where the original $\mathbb{Z}_{4}$ symmetry is reduced to a $\mathbb{Z}_{2}$ symmetry. The Bargmann index $q=1/4$ for the subspace with an even photon number and $q=3/4$ for the subspace with an odd photon number. The zeros of the $G$ function correspond to the exact eigenenergies of the Hamiltonian \eqref{H_origin}.
	
	The vanishing denominator in Eq.~\eqref{fnq} determines the positions of the $n > 0$ poles of the $G$ function, given by	
	\begin{equation}
		E_{n>0}^{(q,\mathrm{pole})} = \frac{2(n+q)\beta}{\gamma^{2}} - \frac{1}{2\gamma^{2}} - \frac{U\Delta}{2},  \label{pole}
	\end{equation}
	while the zeroth pole is instead determined by the zero of the denominator in $\Omega_{0}^{(q)}$, yielding
	\begin{equation}
		E_{0}^{(q,\mathrm{pole})} = \frac{2q\beta}{\gamma} - \frac{\Delta}{2U} - \frac{1}{2\gamma} \left( 1-\frac{\Delta}{U} \right).  \label{pole0}
	\end{equation}
	
	\begin{figure}[tbp]
		\centering
		\includegraphics[width=\linewidth]{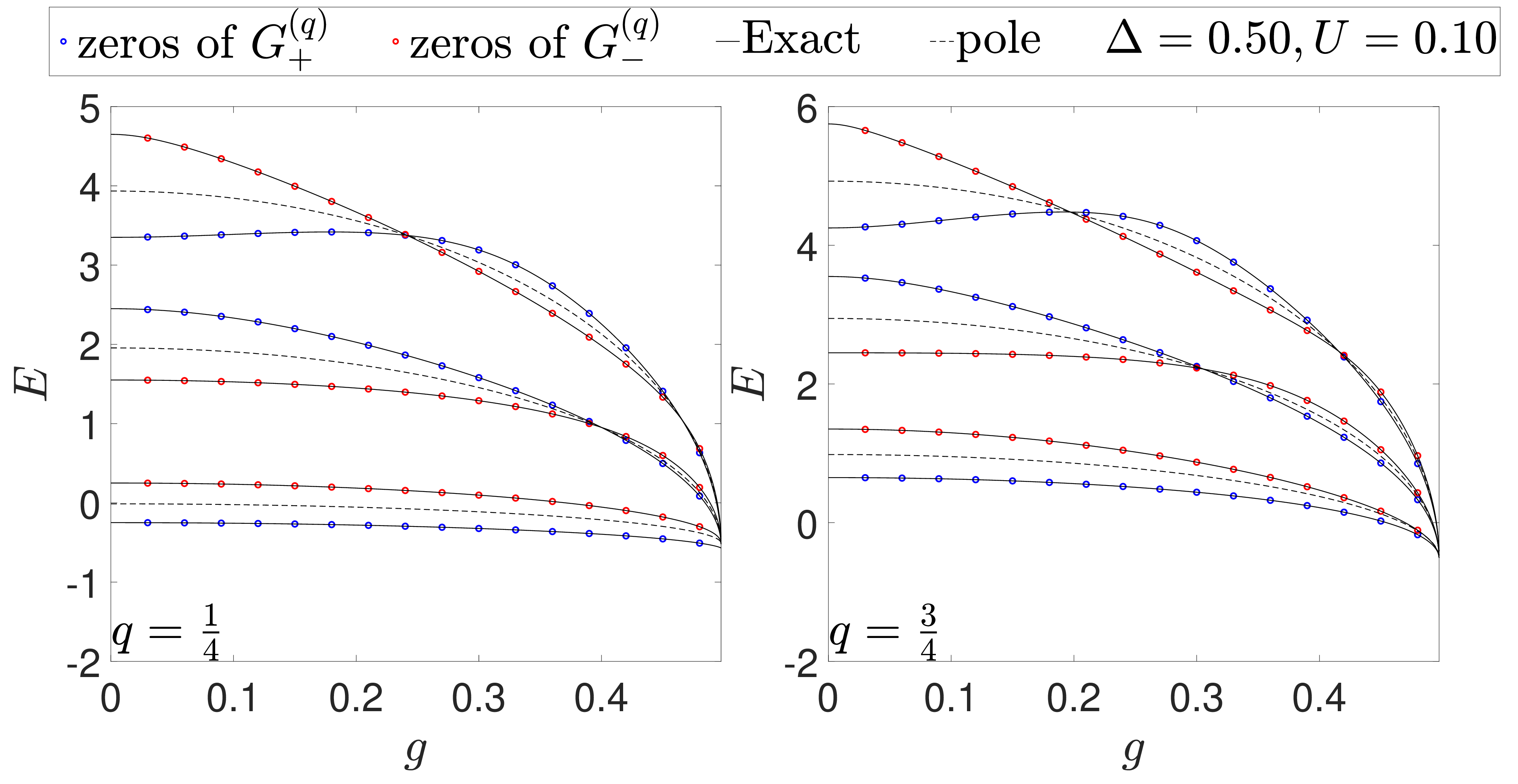}
		\caption{Energy spectra at $\Delta = 0.50$ and $U = 0.10$ as functions of  $g$ for $q= 1/4$ (left panel)  and $q = 3/4$ (right panel). The black solid lines represent eigenvalues obtained through ED, while the dashed lines denote the pole lines $E_n^{(q, \mathrm{pole})}$. Blue circles indicate the zeros of $G_+^{(q)}$, while red ones represent zeros of $G_-^{(q)}$. A spectral collapse occurs at $g_c=\sqrt{1-U^2}/2 \simeq 0.4975$.}
		\label{fig_spectra_G}
	\end{figure}
	
	The energy spectrum is plotted in Fig.~\ref{fig_spectra_G}, obtained either from the zeros of the $G$ function or through numerical exact diagonalization (ED), both of which yield identical results. Along the pole lines, two adjacent energy levels with opposite parity intersect, leading to doubly degenerate eigenenergies at the crossing points. $\beta = 0$, corresponding to $\theta \rightarrow \infty$, determines the maximum coupling strength as 
	\begin{equation}
		g_{c} = \frac{1}{2} \sqrt{1 - U^{2}},
	\end{equation}
	referred to as the critical coupling. At $g=g_c$, all $n>0$ poles collapse to a single value
	\begin{equation}
		E_{n}^{(c)}= - \frac{1}{2\gamma^{2}} - \frac{U\Delta}{2}.
	\end{equation} 
	In the vicinity of $g = g_{c}$, the energy levels become densely distributed. Note that the $G$ function is, however, not well defined exactly at $g_{c}$.	
	
	We begin by performing the asymptotic analysis based on the $G$ function in the regime where the coupling strength approaches $g_{c}$. As shown in Ref.~\cite{li_two-photon_2020}, the condition under which both the denominator and numerator of $\Omega_{n}^{(q)}$ in Eq.~\eqref{enq} simultaneously vanish determines the coupling strength at which the lowest doubly degenerate points occur on each pole line, given by	
	\begin{equation}
		\beta_{c}^{n} = \frac{1-\Delta/U}{4(n+q)},
	\end{equation}
	which requires $\Delta/U \leqslant 1$. Intriguingly, the corresponding pole energies at $\beta_{c}^{n}$ are identical for all $n$
	\begin{equation}
		E_{n}^{\mathrm{cross}} = -\frac{\Delta}{2U} \geqslant  -\frac{1}{2}.
	\end{equation}
	Remarkably, when $\Delta = U$, $\beta_{c}^{n} = 0$, implying that at all these crossing points the coupling strength equals $g_{c}$, and the energies $E_{n}^{\mathrm{cross}}$ are exactly $E_{c} = -1/2$. This implies that, under the condition $\Delta_{c} = U$, all energies collapse to a single value $E_{c} = -1/2$ at the collapse point $g_{c}$. Although the $G$ function is not well defined directly at $g_{c}$, this collapse behavior can still be analytically inferred by extending the pole lines continuously to $g_{c}$. This behavior closely resembles that of the anisotropic tpQRM \cite{li_critical_2025}. We rigorously establish the condition for full spectral collapse based on the identical crossing energy shared by all pole lines. This will be further confirmed through the analysis of the effective Schr\"{o}dinger equation at the collapse point, as discussed in Sec.~\ref{sec_3}.	
	
	\begin{figure}[tbp]
		\includegraphics[width=1.0\linewidth]{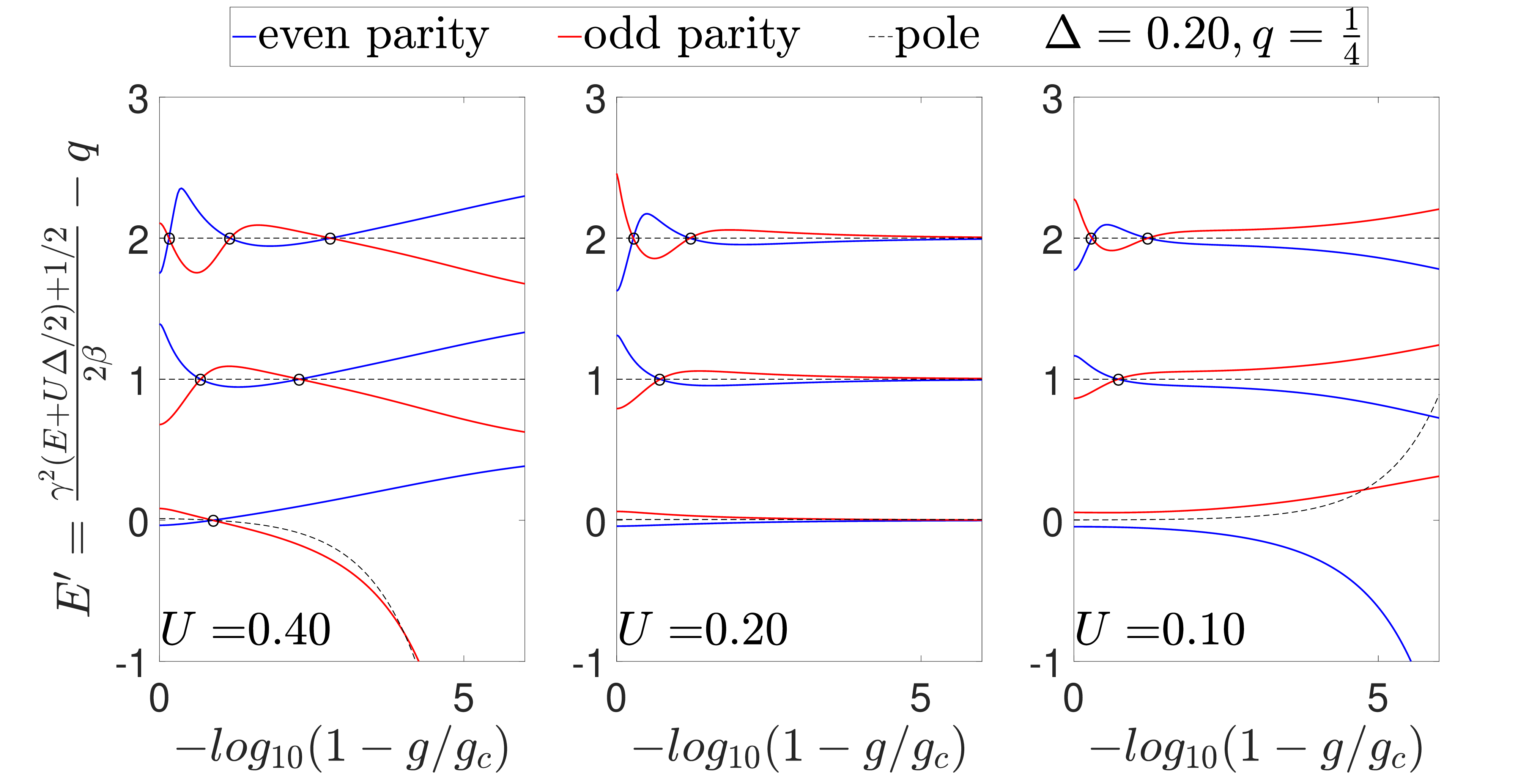}
		\caption{The scaled energy spectra $E^{\prime} = [\gamma^{2} (E+U\Delta/2) + 1/2] /(2\beta) - q$ for $\Delta=0.2$ and $q=1/4$ at $U=0.40$ (left),  $U=0.20$ (middle), and $U=0.10$ (right). The horizontal axis is logarithmically scaled, given by $x=-\log _{10}(1-g/g_{c})$. Blue (red) solid lines indicate energy levels with even (odd) parity, while black dashed lines represent the pole lines. The open circles denote the doubly degenerate points.}
		\label{fig_spectra_trans}
	\end{figure}
	
	To illustrate the complete spectral collapse, energy spectra are presented as a scaled form based on the pole lines, $E^{\prime} = [\gamma^{2} (E+U\Delta/2) + 1/2] /(2\beta) - q$, for three representative qubit-to-Stark ratios: $\Delta/U = 1/2$, $\Delta/U = 1$, and $\Delta/U = 2$, as shown in Fig.~\ref{fig_spectra_trans}. It is evident that only when $\Delta_c = U$, all energy levels will become doubly degenerate at $g = g_{c}$ with $E_c = -1/2$. This indicates a full spectral collapse with no discrete bound states remaining.	
	
	\begin{figure}[tbp]
		\includegraphics[width=1.0\linewidth]{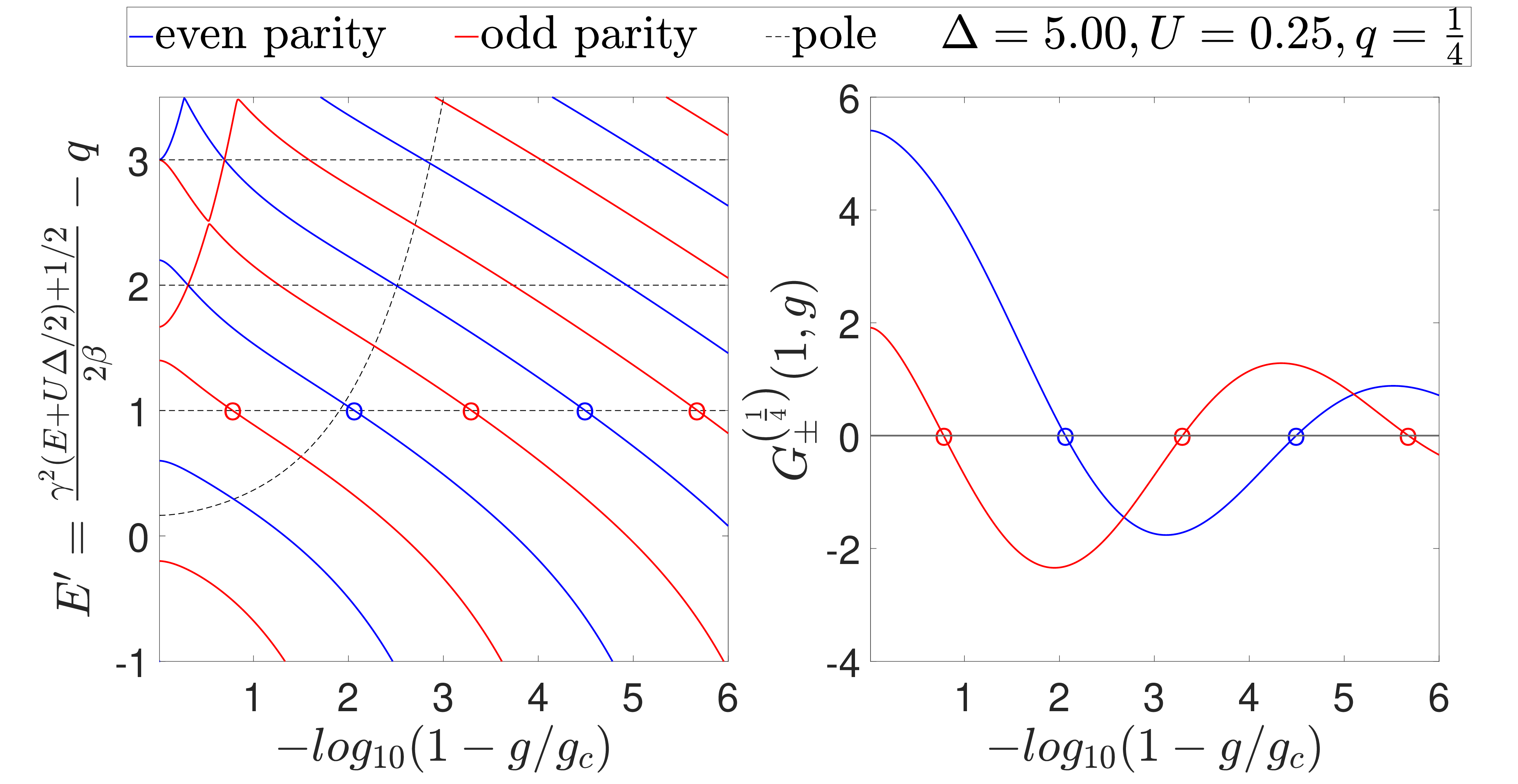}
		\caption{The left panel presents the scaled energy spectra $E^{\prime} = [\gamma^{2} (E+U\Delta/2) + 1/2] /(2\beta) - q$, while the right panel displays the special $G$ function $G_{\pm}^{(q)}(m, g)$ associated with the $m = 1$ pole line. The horizontal axis is logarithmically scaled as $x = -\log_{10}(1 - g / g_c)$. Blue (red) solid lines represent even (odd) parity, and black dashed lines indicate pole positions. The open circles mark the positions of the special nondegenerate points. The parameters are $\Delta = 5.00$, $U = 0.25$, and $q = 1/4$.}
		\label{fig_nond}
	\end{figure}
	
	A natural question arises: if $\Delta/U \neq 1$, what are the characteristics of the energy spectrum at $g=g_{c}$? At the critical coupling, the $G$ function is not well defined, and the ED technique also fails to yield convergent results.	However, the energy spectrum can still be calculated in the coupling regime extremely close to $g_{c}$ using the $G$ function technique. The scaled energy spectrum $E^{\prime}$ is shown in the left panel of Fig.~\ref{fig_nond}. It can be immediately observed that the $n = 0$ pole may lie above the $n > 0$ poles. More interestingly, the energy level may cross the $n = 1$ pole alone and asymptotically fall below all the pole energies. Since all $n >0$ pole energies converge to the same value $E_{n}^{(c)}$ as $g$ approaches $g_{c}$, the corresponding energy levels form discrete bound states.	
	
	The crossing points, denoted by open circles in the left panel of Fig.~\ref{fig_nond}, are referred to as special nondegenerate points, in contrast to the doubly degenerate points shown in Fig.~\ref{fig_spectra_trans}. These nondegenerate points in the tpRSM can be identified using a special $G$ function. If $f_{m-1}^{(q)} = f_{m-2}^{(q)} = 0$ in the three-term recurrence relation \eqref{fnq}, then $f_{n}^{(q)} = 0$ for all $n < m$, indicating a special nondegenerate point where a single energy level intersects the $m$th pole line without degeneracy. With a modified initial condition, the recurrence relations for $e_{n}^{(q)}$ and $f_{n}^{(q)}$ in Eqs.~\eqref{enq} and \eqref{fnq} remain valid. For nondegenerate points on the $m \geqslant  1$ pole lines, the corresponding special $G$ function is defined as:	
	\begin{eqnarray}
		&&G_{\pm}^{(q)}(m,g) = 
		\nonumber \\
		&&\sum_{n=m}^{\infty} \left( e_{n}^{(q)}(g) \mp f_{n}^{(q)}(g) \right) \frac{\left[ 2\left( n+q-\frac{1}{4} \right)\right] !}{2^{n}n!} \tanh^{n}\theta,
		\label{Gexc}
	\end{eqnarray}
	with the initial condition $f_{m}^{(q)}(g) \equiv 1$. For special nondegenerate points on the zeroth pole line, where $\Omega_{0}^{(q)} \rightarrow \infty$, we set $e_{0}^{(q)}(g) \equiv 1$ and $f_{0}^{(q)}(g) \equiv 0$.	In this case, the special $G$ function becomes:	
	\begin{eqnarray}
		&&G_{\pm }^{(q)}(0,g) = 2(q-1/4)! + 
		\nonumber \\
		&&\sum_{n=1}^{\infty} \left[ e_{n}^{(q)}(g) \mp f_{n}^{(q)}(g)\right] \frac{\left[ 2\left( n+q-\frac{1}{4} \right) \right] !}{2^{n}n!} \tanh^{n}\theta.
		\label{Gexc0}
	\end{eqnarray}
	The zeros of these special $G$ functions determine the positions of special nondegenerate points along the pole lines. Importantly, the number of such points is directly related to the number of bound states in the system. As illustrated in Fig.~\ref{fig_nond}, the zeroth pole is energetically separated from the collapse point and lies higher than the other pole lines near $g = g_{c}$. As a result, the number of bound states is effectively determined by the number of special nondegenerate points on the $m = 1$ pole line. The zeros of the special $G$ functions appear at logarithmically spaced intervals along the scaled coupling axis. This behavior closely resembles that observed in the anisotropic tpQRM \cite{li_critical_2025}, suggesting that, under certain parameter regimes, the tpRSM may also support an infinite number of bound states at the collapse point. These states accumulate toward the collapse energy in an exponential fashion.	
	
	We are aware of a recent study on spectral collapse in the anisotropic tpRSM~\cite{yan_analytical_2024}. In that study, only the crossing point associated with the $n = 0$ pole was considered, and the corresponding coupling strength was shifted to $g_{c}$, leading to the claim of a full spectral collapse. However, this argument is incomplete, as the crossing points associated with the $n > 0$ poles were not examined. Without demonstrating the occurrence of all these doubly degenerate states at $g_{c}$, the existence of a full spectral collapse cannot be rigorously established.
	
	\section{Energy spectra at the critical coupling} \label{sec_3}
	
	The $G$ function technique is not applicable exactly at $g_c$. Instead, we employ an alternative representation to investigate the energy spectra of the tpRSM at $g_c$ in this section. The Hamiltonian \eqref{H_origin} can be rewritten in terms of generalized position and momentum operators~\cite{xie_quantum_2019}, with a scaling factor $\kappa$, as:	
	\begin{equation} 
		\kappa x = \frac{a + a^\dagger}{\sqrt{2}}, \quad 
		\kappa^{-1} p = i \frac{a^\dagger - a}{\sqrt{2}}. 
	\end{equation}
	In the $x$ representation of $L^2(\mathbb{R})$ at $g = g_c$, the Hamiltonian becomes:
	\begin{eqnarray} 
		H_{c} &=& \frac{1}{2} \left( \kappa^2 x^2 + \kappa^{-2} p^2 \right) + \frac{1}{2\gamma} \left( \kappa^2 x^2 - \kappa^{-2} p^2 \right) \sigma_z \nonumber \\
		&-& \frac{U}{2} \left[ \kappa^2 x^2 + \kappa^{-2} p^2 - \left( 1 - \frac{\Delta}{U} \right) \right] \sigma_x 
		- \frac{1}{2} \id. 
		\label{H_coll} 
	\end{eqnarray}
	
	To proceed, define the operator $H_0 = H_c + \id/2$ and consider an arbitrary normalized state $\vert \Psi \rangle = [\psi_1 \quad \psi_2]^T \in L^2(\mathbb{R}) \otimes \mathbb{C}^2$. Under the condition $\Delta_c = U$, it can be shown that $H_0$ is a positive semi-definite operator. Explicitly,
	\begin{equation}
		\langle \Psi | H_0 | \Psi \rangle = 
		\langle \phi_1 | \kappa^2 x^2 | \phi_1 \rangle 
		+ \langle \phi_2 | \kappa^{-2} p^2 | \phi_2 \rangle \geqslant 0,
	\end{equation}
	where the auxiliary functions $\vert \phi_1 \rangle$ and $\vert \phi_2 \rangle$ are defined as
	\begin{eqnarray}
		\vert \phi_1 \rangle = \sqrt{\frac{\gamma+1}{2\gamma}} \vert \psi_1 \rangle 
		- \sqrt{\frac{\gamma-1}{2\gamma}} \vert \psi_2 \rangle,
		\nonumber \\
		\vert \phi_2 \rangle = \sqrt{\frac{\gamma+1}{2\gamma}} \vert \psi_2 \rangle 
		- \sqrt{\frac{\gamma-1}{2\gamma}} \vert \psi_1 \rangle.
	\end{eqnarray}
	This confirms that the spectrum of $H_c$ satisfies the inequality $E \geqslant E_c = -1/2$ when $\Delta = \Delta_c$, implying the absence of bound states in this critical case. Thus, we rigorously demonstrate full spectral collapse in the tpRSM if $\Delta = U$, with the same value of $E_c=-1/2$ as in the anisotropic tpQRM \cite{li_critical_2025}. In fact, the only case where $H_0$ is not strictly positive definite corresponds to $x \vert \phi_1 \rangle = p \vert \phi_2 \rangle = 0$, leading to non-normalizable $\vert \Psi \rangle$.
	
	The Schr\"{o}dinger equation $H_c \vert \Psi \rangle = E \vert \Psi \rangle$ for the general state $\vert \Psi \rangle = [\psi_1 \quad \psi_2 ]^T$ can be explicitly written as:
	\begin{eqnarray}
		\left( E + \frac{1}{2} \right) \psi_1 &=& \left[ \frac{\gamma + 1}{2\gamma} \kappa^2 x^2 - \frac{\gamma - 1}{2\gamma} \frac{\partial^2}{\kappa^2 \partial x^2} \right] \psi_1 \nonumber \\
		&-& \frac{U}{2} \left[ \kappa^2 x^2 - \frac{\partial^2}{\kappa^2 \partial x^2} - \left( 1 - \frac{\Delta}{U} \right) \right] \psi_2, \nonumber \\
		\left( E + \frac{1}{2} \right) \psi_2 &=& \left[ \frac{\gamma - 1}{2\gamma} \kappa^2 x^2 
		- \frac{\gamma + 1}{2\gamma} \frac{\partial^2}{\kappa^2 \partial x^2} \right] \psi_2 \nonumber \\
		&-& \frac{U}{2} \left[ \kappa^2 x^2 - \frac{\partial^2}{\kappa^2 \partial x^2} - \left( 1 - \frac{\Delta}{U} \right) \right] \psi_1.
		\label{coll-Schr}
	\end{eqnarray}
	Define a new basis
	\begin{equation}
		\Phi_1 = \sqrt{1 - U}\, (\psi_1 + \psi_2), \quad
		\Phi_2 = \sqrt{1 + U}\, (\psi_1 - \psi_2).
	\end{equation}
	By adding and subtracting the equations in Eq.~\eqref{coll-Schr}, we obtain:
	\begin{eqnarray}
		&&\left( \kappa^2 x^2 - \frac{\partial^2}{\kappa^2 \partial x^2} \right) \Phi_1 
		+ \left( \kappa^2 x^2 + \frac{\partial^2}{\kappa^2 \partial x^2} \right) \Phi_2 \nonumber \\
		&&= \left( \frac{2E + \Delta}{1 - U} + 1 \right) \Phi_1, \nonumber \\
		&&\left( \kappa^2 x^2 - \frac{\partial^2}{\kappa^2 \partial x^2} \right) \Phi_2 
		+ \left( \kappa^2 x^2 + \frac{\partial^2}{\kappa^2 \partial x^2} \right) \Phi_1 
		\nonumber \\
		&&= \left( \frac{2E - \Delta}{1 + U} + 1 \right) \Phi_2.
		\label{coll}
	\end{eqnarray}
	
	Next, define the symmetric and antisymmetric combinations:
	\begin{equation}
		\varphi_1 = \Phi_1 + \Phi_2, \quad \varphi_2 = \Phi_1 - \Phi_2,
	\end{equation}
	and perform further linear combinations of Eq.~\eqref{coll} to decouple the system. This yields:
	\begin{eqnarray}
		- \frac{\partial^2 \varphi_2}{\kappa^2 \partial x^2} &=& \gamma^2 \left[ E-E_n^{(c)} \right] \varphi_2 
		+ \gamma^2 U \left( E + \frac{\Delta}{2U} \right) \varphi_1, \nonumber \\
		\kappa^2 x^2 \varphi_1 &=& \gamma^2 \left[ E-E_n^{(c)} \right] \varphi_1 + \gamma^2 U \left( E + \frac{\Delta}{2U} \right) \varphi_2.
	\end{eqnarray}
	Solving for $\varphi_2$ leads to:
	\begin{equation}
		- \frac{\partial^2 \varphi_2}{\kappa^2 \partial x^2} = \gamma^2 \left[ E-E_n^{(c)} \right] \varphi_2 + \frac{ \left[ \gamma^2 U \left( E + \frac{\Delta}{U} \right) \right]^2 }{ \kappa^2 x^2 - \gamma^2 \left[ E-E_n^{(c)} \right] } \varphi_2.
		\label{coll_xorigin}
	\end{equation}
	To simplify the equation, we choose $\kappa^2$ such that:
	\begin{equation}
		0 < \kappa^2 = -\gamma^2 \left[ E-E_n^{(c)} \right] = -\frac{1}{2} - \frac{E + \frac{\Delta U}{2} }{1-U^2}, \label{kappa2}
	\end{equation}
	which ensures that we are probing the bound state regime, i.e., $E < E_n^{(c)}$. Under this condition, Eq.~\eqref{coll_xorigin} reduces to a Schr\"{o}dinger-type equation:
	\begin{equation}
		- \frac{\partial^2}{\partial x^2} \varphi_2 + V(x) \varphi_2 = -\kappa^4 \varphi_2,
		\label{coll_x}
	\end{equation}
	where the effective potential is given by
	\begin{equation}
		V(x) = - U^2 \cdot \frac{ \left[ \kappa^2 + \frac{1}{2} \left( 1 - \frac{\Delta}{U} \right) \right]^2 }{ x^2 + 1 }.
		\label{V}
	\end{equation}
	
	Equation~\eqref{coll_x} describes a one-dimensional particle subject to a smooth attractive potential well $V(x)$, with energy eigenvalue $-\kappa^4$. This formulation enables the application of standard spectral analysis to determine the existence of bound states. Meanwhile, when $\Delta=\Delta_c$, Eq.~\eqref{coll_x} becomes
	\begin{equation}
		- \frac{\partial^2}{\partial x^2} \varphi_2 = - \kappa^4 \left( 1-\frac{U^2}{x^2 + 1} \right) \varphi_2.
	\end{equation}
	Since $U^2 < 1$, both the operators $p^2 = -\partial^2/(\partial x^2)$ and $1-U^2/(x^2 + 1)$ are positive semi-definite. The equation cannot support a negative eigenvalue, and thus, no bound states exist. For the general case where $\Delta / U \neq 1$, to discuss qualitatively the number of bound states, we may use Faddeev's criterion \cite{drazin1989}: The potential $V_1(x)$ allows for at most a finite number of bound states if
		\begin{equation}
			I_1=\int_{-\infty}^\infty\textrm{d}x |V^{(-)}_1 (x)|(1+|x|) < \infty,
			\label{II}
		\end{equation}
		where $V_1^{(-)}(x)=0$ whenever $V_1(x)>0$ and coincides with $V_1(x)$ elsewhere. Obviously, here $I_1$ is always logarithmically divergent, and therefore, an infinite number of bound states exist.
	
	\begin{figure}[tbp] 
		\includegraphics[width=1.0\linewidth]{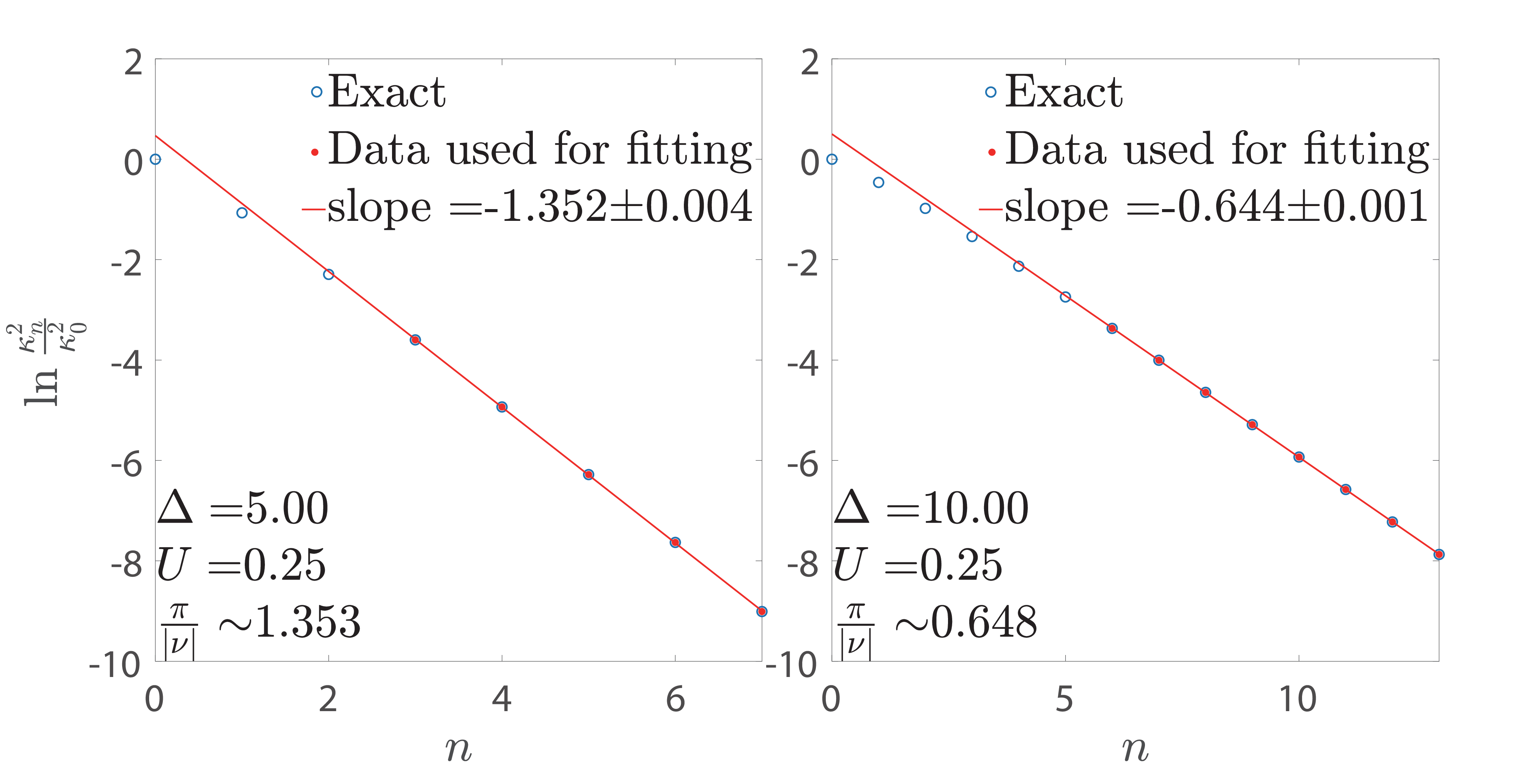} 
		\caption{ $\ln (\kappa_n^2 / \kappa_0^2)$ for the first several bound states at the critical coupling, $g = g_c$, $U = 0.25$, and $q = 1/4$, for $\Delta = 5.00$ (left) and $\Delta = 10.00$ (right). The open blue circles represent all data obtained via numerical ED, while the red filled circles denote the data used for fitting, which approaches the threshold energy $E_n^{(c)}$. The red line indicates a linear fit.}
		\label{fig_en_ln} 
	\end{figure}
	
	When the value of $\Delta$ is much greater than $U$, it can be seen from \autoref{fig_nond} that the bound states may approach the threshold energy exponentially. In this case, the problem can be mapped to a one-dimensional inverse-square potential well to obtain an approximate result for the bound energy levels. By using matching conditions with a constant potential in the central region, Nguyen and Marsiglio \cite{nguyen_numerical_2020} provided an analytical approximation for the bound energy levels in the inverse-square potential well. As the effective potential \autoref{V} behaves asymptotically as $\propto x^{-2}$ when $x \to \infty$ and is bounded from below at $x = 0$, the difference has a small effect on the results. By introducing the variable substitutions $\rho = \kappa^2 x$ and $\varphi_2(x) = \sqrt{\rho}\, y(\rho)$, the differential equation can be transformed into the modified Bessel equation when $x \to \infty$:
		\begin{equation}
			\rho^2 \frac{d^2 y}{d\rho^2} + \rho \frac{d y}{d\rho} - \left( \rho^2 + \nu^2 \right) y = 0,
		\end{equation}
		where
		\begin{equation}
			\nu^2 = - U^2 \left[ \kappa^2 + \frac{1}{2} \left( 1 - \frac{\Delta}{U} \right) \right]^2 + \frac{1}{4}.
		\end{equation}
		The eigenvalues of the even-parity wave function ($q=1/4$) satisfy:
		\begin{equation}
			-\kappa_n^4 = -\kappa_0^4 \exp \left[ -\frac{2 \pi n}{|\nu|} \right], \quad n = 1, 2, 3, \dots
		\end{equation}
		where $-\kappa_0^4$ is the lowest eigenvalue. Combining with \autoref{kappa2}, the bound state energies in the two-photon quantum Rabi-Stark model can be expressed as:
		\begin{equation}
			\frac{\kappa_n^2}{\kappa_0^2} = \frac{E_n - E_n^{(c)}}{E_0 - E_n^{(c)}} = \exp \left( -\frac{\pi n}{|\nu|} \right),
		\end{equation}
		where $E_0$ represents the ground state energy. Clearly, the bound state energies approach the threshold energy $E_n^{(c)}$ exponentially, with a decay rate of $\pi / |\nu|$, and can be approximated as $\nu^2 \sim [1 - (\Delta - U)^2] / 4$, because near the threshold energy, $\kappa^2 \to 0$. As shown in \autoref{fig_en_ln}, this approximation agrees well with the numerical results for highly excited bound states (large $n$), where these energy levels approach the threshold energy exponentially, indicating that the inverse-square potential well is highly reliable in describing this behavior.  
	
	\section{Quantum criticality}
	
	As discussed above, the emergent nonlinear Stark interaction induces novel and exotic physical properties in the tpRSM. In particular, when $\Delta = U$, as the coupling strength increases, the energy levels coalesce to a common value at the critical coupling strength $g_c$. This behavior indicates that the energy gap between the first excited state and the ground state closes at a finite frequency ratio $\Delta / \omega$ in the tpRSM, similar to its one-photon counterpart \cite{xie_quantum_2019}.	
	
	\begin{figure}[tbp]
		\includegraphics[width=1.0\linewidth]{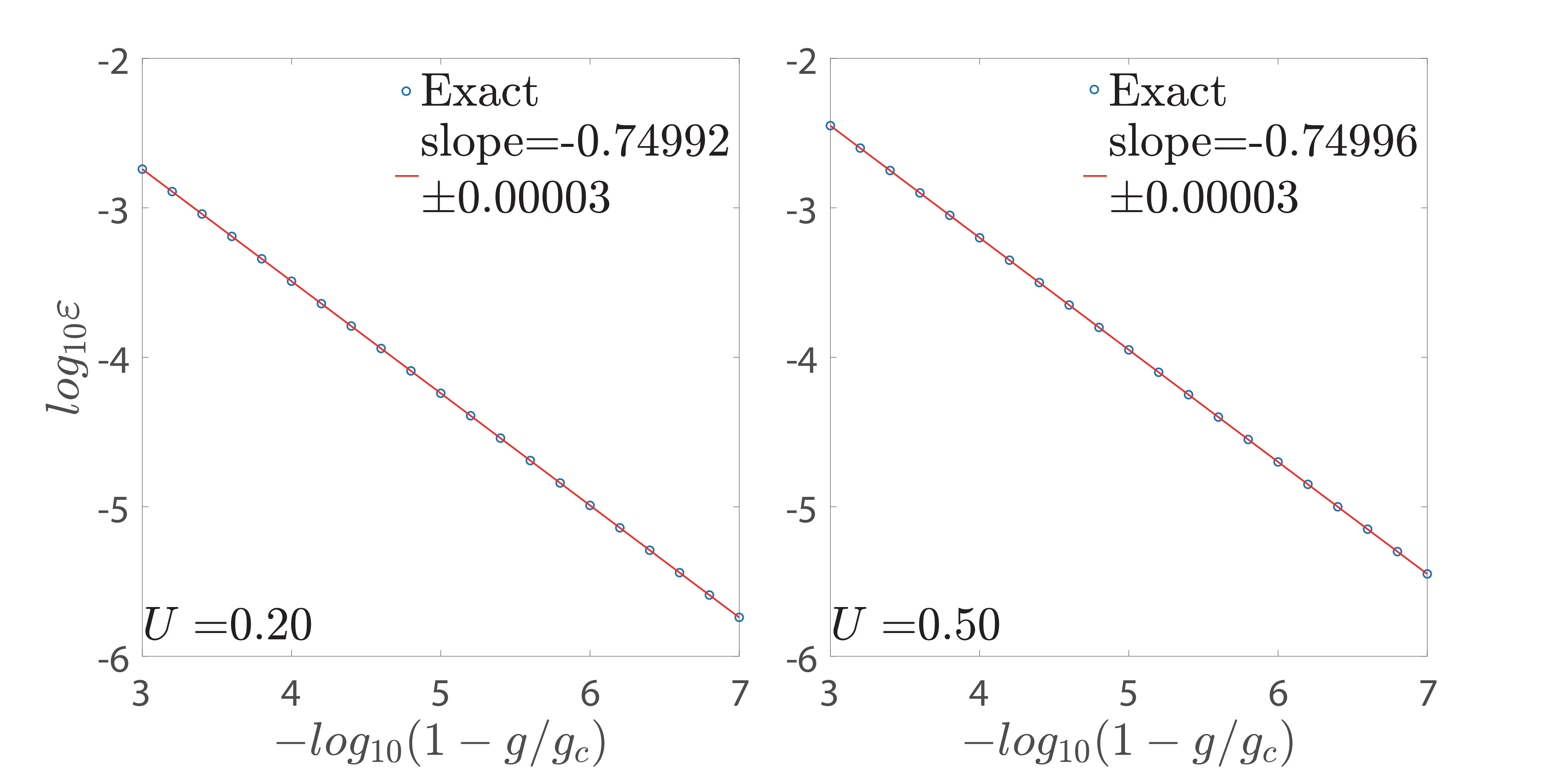}
		\caption{Logarithmically scaled energy gap under the condition $\Delta = \Delta_c$ for $U=0.20$ (left) and $U = 0.50$ (right). The horizontal axis is logarithmically scaled as $x=-\mathrm{log}_{10} (1-g/g_c)$; Open circles represent results obtained via numerical ED, while the red line denotes a linear fit.}
		\label{fig_energygap_g}
	\end{figure}
	
	We examine the energy gap, $\varepsilon (g\rightarrow g_{c}) \sim \left\vert
	g-g_{c} \right\vert ^{z\nu_x }$ under $\Delta =U$, where $\nu_x$ is the critical exponent, and $z$ is the dynamical exponent. The product $z\nu_x$ is the energy gap exponent. As shown in Fig.~\ref{fig_energygap_g}, the exponent for the coupling-induced gap closure is $3/4$, independent of the value of $U$. This gap exponent is smaller than the value 2 observed in the one-photon RSM \cite{xie_quantum_2019} and greater than 1/2 in the QRM \cite{ashhab_superradiance_2013, hwang_quantum_2015}, indicating a distinct universality class.
	
	The QPT occurs in the quantum Rabi model at an infinite ratio of the qubit to cavity frequencies $\Delta/\omega$,  whereas both the tpRSM and one-photon RSM experience the QPT at finite ratios of $\Delta/\omega$. In the one-photon RSM, the QPT emerges only at the Stark coupling strength $U/\omega=1$ \cite{chen_quantum_2020}, whereas the tpRSM can experience a QPT at any finite $U/\omega < 1$ as long as $\Delta = U$. The gap exponent $z\nu_x = 3/4$ in the tpRSM, which lies between those of the quantum Rabi model and the one-photon RSM, reflects a nontrivial critical behavior governed by the interplay between two-photon interactions and the nonlinear Stark term.
	
	\begin{figure}[tbp]
		\includegraphics[width=1.0\linewidth]{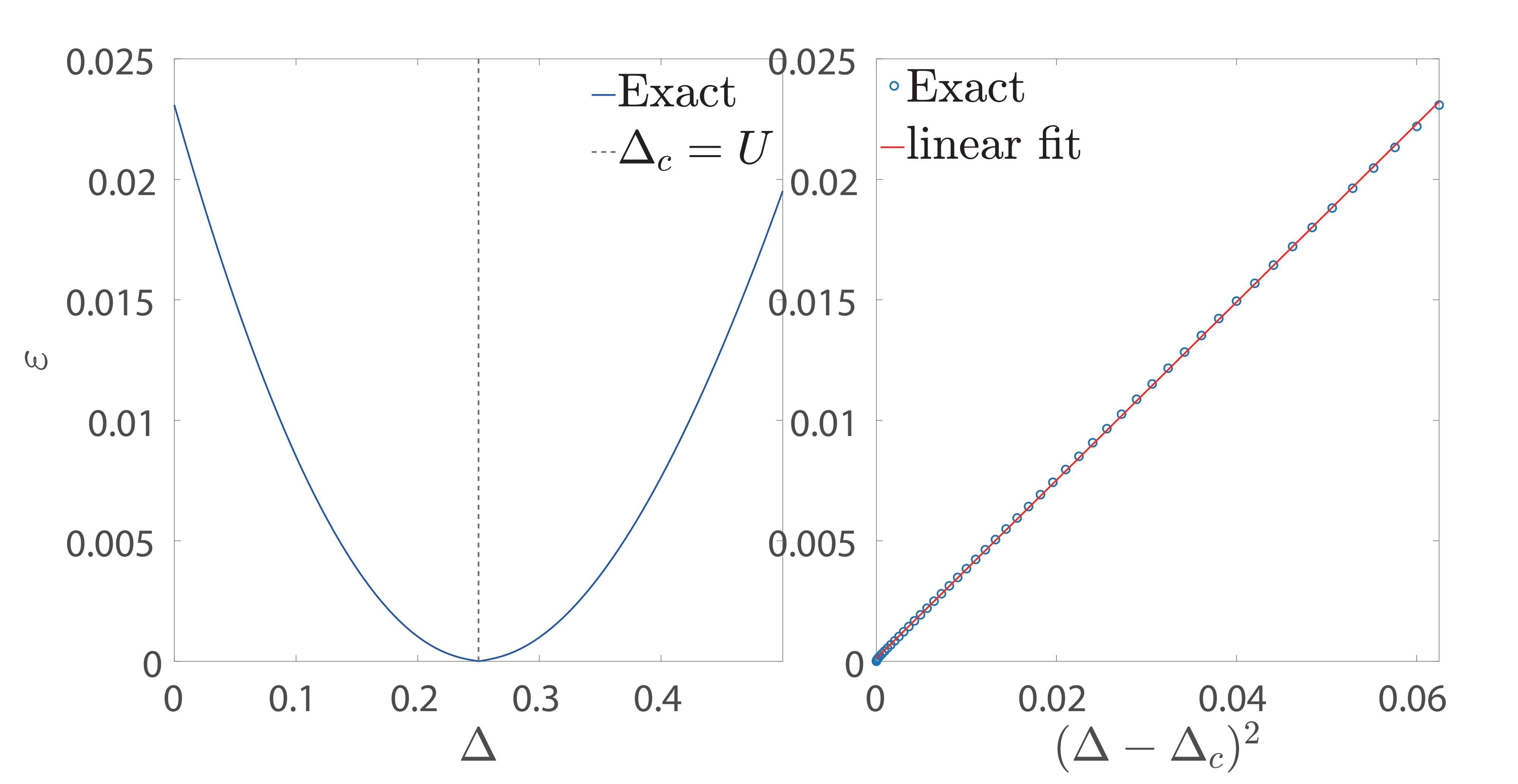}
		\caption{Left: Energy gap at $U=0.25$ and $g=g_{c}$ as a function of $\Delta$ by numerical ED, and the black dashed line marks the critical value $\Delta_{c}=U$. Right: Replot of the gap curves from the left panel as a function of $(\Delta - \Delta_c)^2$ in the vicinity of $\Delta < \Delta_c$ (open circles). The red line shows a linear fit.}
		\label{fig_energygap}
	\end{figure}
	
	For finite $U$, we can also analyze the energy gap $\varepsilon (\Delta \to U)$ as a function of $\Delta$ at the critical coupling $g_{c} = \sqrt{1 - U^{2}}/2$.	The energy gap for $U = 0.25$ calculated via ED is shown in the left panel of Fig.~\ref{fig_energygap}. The gap curve exhibits an approximately quadratic dependence and reaches zero precisely at $\Delta_c = U$. This observation confirms the absence of bound states at the spectral collapse point.	To further elucidate this behavior, we plot the energy gap $\varepsilon$ as a function of $(\Delta - \Delta_c)^{2}$ in the right panel, focusing on the regime $\Delta < \Delta_c = U$.	A linear fit yields excellent agreement with the numerical data. This high degree of accuracy strongly supports the conclusion that the energy gap closes quadratically as $\Delta \to U$ from below. Such quadratic scaling is a hallmark of a second order QPT, indicating that the system undergoes a continuous phase transition at the critical point. The quadratic dependence typically arises from symmetry-protected energy crossings or perturbative energy shifts. This indicates that the gap closes in a parabolic manner due to second order effects in perturbation theory.	Consequently, this behavior suggests a relatively smooth and weakly critical transition, potentially governed by a simpler effective Hamiltonian near $\Delta_c$.	
	
	The phase transition occurs in the thermodynamic limit, characterized by the emergence of singularities in certain physical observables. Despite having few degrees of freedom in the QRM, the effective system size can be defined as $L_\mathrm{R} = \Delta / \omega$ \cite{hwang_quantum_2015} to gain insight into the critical behavior of the phase transitions.	The singularity appears only in the limit of  $\Delta /\omega \rightarrow \infty $. In the present tpRSM, since the energy gap closes only at $U = \Delta$, we define the system size here as:
	\begin{equation}
		L=\frac{1}{\left \vert \Delta/U-1 \right \vert}.
	\end{equation}
	For the tpRSM with finite $U < 1$, as we increase the qubit frequency $\Delta$ from below, the energy gap at $g = g_c$ generally opens. Once $\Delta$ reaches $U$, the gap closes. As $\Delta$ exceeds $U$, the energy gap opens again. With this definition, $\Delta = U$ corresponds to the thermodynamic limit.
	
	In the absence of nonlinear Stark coupling, for any finite qubit frequency, the spectrum at $g_c = 1/2$ consists of both discrete and continuous components. The ground state is always separated from the continuum by a finite excitation gap, ruling out a QPT in the usual sense \cite{duan_two-photon_2016}. Specifically, the full spectrum collapse occurs at $\Delta = 0$ when $U = 0$.	In this case, the model decouples into the upper and lower states of the qubit. The Hamiltonian \eqref{H_origin} becomes
	\begin{equation}
		H_\mathrm{R} = \omega a^\dagger a + g \sigma_z \left[ a^2 + (a^\dagger)^2 \right]
		= \begin{bmatrix}
			\beta _0 A^\dagger A - \frac{1}{2} & 0\\
			0	& \beta _0 B^\dagger B - \frac{1}{2}
		\end{bmatrix}
		\nonumber \\
	\end{equation}
	where 
	\begin{equation}
		A = a \cosh \theta_0 + a^\dagger \sinh \theta_0,
		\quad
		B = a \cosh \theta_0 - a^\dagger \sinh \theta_0,
	\end{equation}
	with $\tanh \theta_0 = \sqrt{(1-\beta_0) / (1+\beta_0)}$ and $\beta_0 = \sqrt{1-4g^2}$. As a result, in the $q = 1/4$ or $q = 3/4$ subspace, the eigenvalues are always doubly degenerate and equal to $2(n+q)\beta_0 - 1/2$ for $g<g_c=1/2$. The energy gap between the ground state and the first excited state remains zero.
	
	\section{Conclusion}
	
	In this work, we have conducted a comprehensive analytical and numerical study of the tpRSM. By employing the $G$ function method within the $sl_2(\mathbb{R})$ Lie algebra framework, we derived an exact expression for the pole structure and established the critical coupling strength $g_c$ for spectral collapse. Our analysis reveals that, unlike in the tpQRM, the Stark term induces an isolated zeroth pole and fundamentally modifies the spectrum’s collapse behavior.
	
	We demonstrate that the existence of discrete bound states at the collapse point depends sensitively on the qubit splitting parameter $\Delta$. In particular, we find that only when $\Delta = \Delta_c = U$, all bound states vanish and the spectrum becomes fully continuous. For $\Delta \neq U$, an infinite number of bound states emerges. Numerical analysis of the special $G$ function and asymptotic analysis of the transformed Schr\"{o}dinger-type equation with an effective inverse square potential, reveal logarithmic accumulation of these bound energy levels near the collapse threshold.
	
	Our analysis reveals that, at the collapse point, the energy gap closes following a quadratic scaling law with respect to $\Delta$, indicating a second order quantum phase transition. We extract the universal energy gap exponent $z\nu_x = 3/4$ for $g$, which places the model in a distinct universality class. More interestingly, the critical coupling strength $g_c=\sqrt{1-U^2}/2$ decreases with the Stark coupling $U$, thereby facilitating the experimental feasibility of the exotic spectral properties and the criticality of this model at weaker coupling. Our findings not only deepen the understanding of spectral collapse phenomena and quantum criticality but also open new possibilities for controlling quantum phase transitions through nonlinear interactions in light-matter coupling systems.		
	
	\section*{Acknowledgments}
	
	We acknowledge useful discussions with Daniel Braak.  This work is supported by the National Key R$\&$D Program of China under Grant No. 2024YFA1408900.

	\section*{Appendix: Derivation of $G$ function} \label{sec_appendix}
	
	Begin by applying the Bogoliubov (or squeezing) transformation,
	\begin{equation}
		S(\theta) = \exp \left[ \frac{\theta}{2} \left( (a^{\dagger})^{2} - a^{2} \right) \right],
	\end{equation}
	and make use of the following identities:
	\begin{eqnarray}
		S(\theta)\, a\, S^\dagger(\theta) = a \cosh \theta - a^\dagger \sinh \theta, \nonumber \\
		S(\theta)\, a^\dagger\, S^\dagger(\theta) = a^\dagger \cosh \theta - a \sinh \theta,
	\end{eqnarray}
	where the hyperbolic functions are given by
	\begin{equation}
		\cosh \theta = \sqrt{\frac{1 + \beta}{2\beta}}, \quad
		\sinh \theta = \sqrt{\frac{1 - \beta}{2\beta}},
	\end{equation}
	with $\beta = \sqrt{1 - 4\gamma^2 g^2}$. The transformed Hamiltonian $H_S = S(\theta) H S^\dagger(\theta)$ takes a matrix form with elements:
	\begin{eqnarray}
		H_{11} &=& \frac{1 - 4\gamma g^2}{\beta} \left( a^\dagger a + \frac{1}{2} \right)
		- \frac{\gamma - 1}{\beta} g \left[ a^2 + (a^\dagger)^2 \right] - \frac{1}{2}, \nonumber \\
		H_{12} &=& H_{21} = - \frac{\Delta}{2} 
		- U \left[ \frac{a^\dagger a + \frac{1}{2}}{\beta} 
		- \frac{\gamma}{\beta} g \left( a^2 + (a^\dagger)^2 \right) - \frac{1}{2} \right], \nonumber \\
		H_{22} &=& \frac{1 + 4\gamma g^2}{\beta} \left( a^\dagger a + \frac{1}{2} \right)
		- \frac{\gamma + 1}{\beta} g \left[ a^2 + (a^\dagger)^2 \right] - \frac{1}{2}.
		\nonumber \\
		\label{H_S}
	\end{eqnarray}
	
	A well known presentation of the Lie algebra $sl_2(\mathbb{R})$ is given by
	\begin{equation}
		K_{0} = \frac{1}{2} \left( a^{\dagger}a + \frac{1}{2} \right), \quad 
		K_{+} = \frac{(a^{\dagger})^{2}}{2}, \quad 
		K_{-} = \frac{a^{2}}{2}, \nonumber
	\end{equation}
	with the commutation relations
	\begin{equation}
		\left[ K_{0}, K_{\pm} \right] = \pm K_{\pm}, \quad 
		\left[ K_{+}, K_{-} \right] = -2K_{0}.
	\end{equation}
	The Hamiltonian \eqref{H_S} can be expressed in terms of these $sl_2(\mathbb{R})$ generators. The Hilbert space $L^2(\mathbb{R})$ decomposes into two irreducible subspaces of $sl_2(\mathbb{R})$, spanned respectively by the vectors $\{(a^\dagger)^n |0\rangle\}$ with even and odd $n$. These subspaces are characterized by the Bargmann index $q$, defined as the eigenvalue of $K_0$ acting on the corresponding vacuum (lowest weight) state:
	\begin{equation}
		K_0 |q, 0 \rangle = q\, |q, 0 \rangle, \quad 
		K_- |q, 0 \rangle = 0.
	\end{equation}
	For the even subspace $\mathcal{H}_{\frac{1}{4}} = \left\{ (a^{\dagger})^n |0\rangle , n=0,2,4,\ldots \right\}$, the Bargmann index is $q = 1/4$; for the odd subspace $\mathcal{H}_{\frac{3}{4}} = \left\{ (a^{\dagger})^n |0\rangle , n=1,3,5,\ldots \right\}$, we have $q = 3/4$. The basis states can be written as
	\begin{eqnarray}
		|q, n \rangle &=& \left| 2 \left( q + n - \frac{1}{4} \right) \right\rangle 
		= \frac{(a^{\dagger})^{2(q + n - \frac{1}{4})}}{\sqrt{ \left[ 2(q + n - \frac{1}{4}) \right]! }} |0\rangle, \nonumber \\
		K_0 |q, n \rangle &=& (q + n) |q, n \rangle.
	\end{eqnarray}
	
	Using these generators of the $sl_2(\mathbb{R})$ Lie algebra, the eigenfunctions of the Hamiltonian $H_S$ can be expanded as
	\begin{equation} 
		\left\vert \psi^{(q)} \right\rangle = \begin{bmatrix} 
			\sum_{n=0}^{\infty} \sqrt{ \left[ 2\left( n+q-\frac{1}{4} \right) \right]! }\, e_{n}^{(q)} \left\vert q,n\right\rangle \\ 
			\sum_{n=0}^{\infty} \sqrt{ \left[ 2\left( n+q-\frac{1}{4} \right) \right]! }\, f_{n}^{(q)} \left\vert q,n \right\rangle 
		\end{bmatrix},
		\label{psi_tpRS_q} 
	\end{equation}
	where $e_{n}^{(q)}$ and $f_{n}^{(q)}$ are expansion coefficients. Substituting this into the Schr\"{o}dinger equation $H_S \left\vert \psi^{(q)} \right\rangle = E \left\vert \psi^{(q)} \right\rangle$ and projecting onto $\left\vert q,n \right\rangle$ yields the following coupled equations:
	\begin{widetext}
	\begin{eqnarray} 
		\left[ 2 \frac{1 - 4\gamma g^2}{\beta} (n+q) - \frac{1}{2} - E \right] e_n^{(q)} - \frac{\gamma - 1}{\beta} g \Lambda_n^{(q)} - \left[ \frac{\Delta}{2} + \frac{2U}{\beta} (n+q) - \frac{U}{2} \right] f_n^{(q)} + \frac{U \gamma}{\beta} g F_n^{(q)} = 0, \nonumber \\
		\left[ 2 \frac{1 + 4\gamma g^2}{\beta} (n+q) - \frac{1}{2} - E \right] f_n^{(q)} - \frac{\gamma + 1}{\beta} g F_n^{(q)}  - \left[ \frac{\Delta}{2} + \frac{2U}{\beta} (n+q) - \frac{U}{2} \right] e_n^{(q)} + \frac{U \gamma}{\beta} g \Lambda_n^{(q)} = 0,
		\nonumber \\ \label{eqna} 
	\end{eqnarray}
	where
	\begin{equation}
		\Lambda_n^{(q)} = e_{n-1}^{(q)} + 4 \left( n+q+\frac{1}{4} \right) \left( n+q+\frac{3}{4} \right) e_{n+1}^{(q)}, \quad
		F_n^{(q)} = f_{n-1}^{(q)} + 4 \left( n+q+\frac{1}{4} \right) \left( n+q+\frac{3}{4} \right) f_{n+1}^{(q)}.
	\end{equation}
	By imposing the condition
	\begin{equation}
		\frac{\gamma - 1}{\beta} \cdot \frac{\gamma + 1}{\beta} = \left( \frac{U \gamma}{\beta} \right)^2 \quad \Rightarrow \quad \gamma = \frac{1}{\sqrt{1 - U^2}},
	\end{equation}
	the offdiagonal terms $\Lambda_n^{(q)}$ and $F_n^{(q)}$ are eliminated, leading to a one-to-one correspondence between $e_n^{(q)}$ and $f_n^{(q)}$:
	\begin{equation}
		e_{n}^{(q)} = \frac{\frac{\Delta}{2} + \frac{2U}{\beta}(n+q) - \frac{U}{2} - \frac{U\gamma}{\gamma+1} \left[ 2\frac{1+4\gamma g^{2}}{\beta}(n+q) - \frac{1}{2} - E \right]}{2\frac{1-4\gamma g^{2}}{\beta}(n+q) - \frac{1}{2} - E - \frac{U\gamma}{\gamma +1} \left[ \frac{\Delta}{2} + \frac{2U}{\beta}(n+q) - \frac{U}{2} \right]} f_{n}^{(q)} = \Omega_{n}^{(q)} f_{n}^{(q)},  \label{enq_a}
	\end{equation}
	Substituting Eq.~\eqref{enq_a} into Eq.~\eqref{eqna} yields a recurrence relation for $f_n^{(q)}$:
	\begin{equation}
		f_{n+1}^{(q)} = \frac{\left\{ 2(1+4\gamma g^{2})(n+q) - \beta \left( \frac{1}{2} + E \right) -\left[ \beta \frac{\Delta -U}{2} + 2U(n+q) \right] \Omega_{n}^{(q)} \right\} f_{n}^{(q)} - \left[ \gamma + 1 - U \gamma \Omega_{n-1}^{(q)} \right] g f_{n-1}^{(q)}}{4g \left( n+q+\frac{1}{4} \right) \left( n+q+\frac{3}{4} \right) \left[ \gamma + 1 - U \gamma \Omega_{n+1}^{(q)} \right] }. 
	\end{equation}
	\end{widetext}
	Transforming back to the original Hamiltonian, the eigenfunctions become $\left| \Psi^{(q)} \right\rangle = S(-\theta) \left| \psi^{(q)} \right\rangle$. If $\left| \Psi^{(q)} \right\rangle$ is an eigenstate of the conserved parity operator $\Pi$ with nondegenerate energy, i.e. $\Pi \left| \Psi^{(q)} \right\rangle \propto \left| \Psi^{(q)} \right\rangle$, then
	\begin{equation} 
		\left( \sigma_x\, e^{i \frac{\pi}{2} a^\dagger a} \right) S(-\theta) \left| \psi^{(q)} \right\rangle \propto S(-\theta) \left| \psi^{(q)} \right\rangle.
	\end{equation}
	This implies the proportionality:
	\begin{eqnarray}
		S(\theta) \begin{bmatrix} 
			\sum_{n=0}^{\infty} (-1)^n \sqrt{ \left[ 2(n+q-\frac{1}{4}) \right]! } f_n^{(q)} \left| q,n \right\rangle \\ 
			\sum_{n=0}^{\infty} (-1)^n \sqrt{ \left[ 2(n+q-\frac{1}{4}) \right]! } e_n^{(q)} \left| q,n \right\rangle 
		\end{bmatrix} \propto 
		\nonumber \\
		S(-\theta) \begin{bmatrix} 
			\sum_{n=0}^{\infty} \sqrt{ \left[ 2(n+q-\frac{1}{4}) \right]! } e_n^{(q)} \left| q,n \right\rangle \\ 
			\sum_{n=0}^{\infty} \sqrt{ \left[ 2(n+q-\frac{1}{4}) \right]! } f_n^{(q)} \left| q,n \right\rangle 
		\end{bmatrix}.
		\label{pro_relation}
	\end{eqnarray}
	With the use of
	\begin{equation} 
		S(\theta) \left| q,0 \right\rangle \propto \sum_{n=0}^{\infty} \frac{ \sqrt{ \left[ 2(n+q-\frac{1}{4}) \right]! } }{2^n n!} \tanh^n \theta \left| q,n \right\rangle,
	\end{equation}
	projecting both sides of Eq.~\eqref{pro_relation} onto the vacuum state $\left| q,0 \right\rangle$ yields the $G$ function:
	\begin{equation}
		G_{\pm}^{(q)}(E) = \sum_{n=0}^{\infty} \left( e_n^{(q)} \mp f_n^{(q)} \right) \frac{ \left[ 2(n+q-\frac{1}{4}) \right]! }{2^n n!} \tanh^n \theta,
	\end{equation}
	with the initial condition $f_0^{(q)} \equiv 1$.

\end{document}